# The Disruption Index Measures Displacement Between a Paper and its Most Cited Reference

**Authors:** Yiling Lin[1†], Linzhuo Li[2†*], Lingfei Wu[1*]

**Affiliations:**

[1] School of Computing and Information, University of Pittsburgh, Pittsburgh, PA 15260

[2] Department of Sociology, Zhejiang University, Hangzhou 310058, China

†These authors contributed equally as co-first authors.

*Corresponding authors. E-mail: linzhuoli@zju.edu.cn (L.L.) and liw105@pitt.edu (W.L.)

## Abstract

Initially developed to capture technical innovation and later adapted to identify scientific breakthroughs, the Disruption Index (D-index) offers the first quantitative framework for analyzing transformative research. Despite its promise, prior studies have struggled to clarify its theoretical foundations, raising concerns about potential bias. Here, we show that—contrary to the common belief that the D-index measures absolute innovation—it captures *relative* innovation: a paper's ability to displace its most-cited reference. In this way, the D-index reflects scientific progress as the replacement of older answers with newer ones to the same fundamental question—much like light bulbs replacing candles. We support this insight through mathematical analysis, expert surveys, and large-scale bibliometric evidence. To facilitate replication, validation, and broader use, we release a dataset of D-index values for 49 million journal articles (1800–2024) based on OpenAlex.

## 1. Introduction

Scientific breakthroughs are the engine of progress, driving technological advancements, economic growth, and societal transformation. From the development of mRNA-based COVID-19 vaccines to the rise of artificial intelligence (AI) and progress in quantum computing, transformative discoveries have reshaped industries and redefined global competitiveness. Recognizing the profound impact of such breakthroughs, national agencies and global alliances are increasingly focused on securing leadership in science and technology, making the ability to identify and support emerging research frontiers a strategic priority.

However, for decades, decision-makers have relied on citation-based indicators—such as journal impact factors and the h-index—to identify high-impact research. These metrics often reflect popularity rather than innovation, shaping hiring, promotion, and funding decisions in ways that favor incremental research from established scholars (Franzoni et al., 2022). This evaluation system creates incentives for researchers to maximize publication counts on familiar topics rather than pursue high-risk, high-reward discoveries. As a result, there is growing concern about scientific stagnation (Chu & Evans, 2021), where transformative ideas struggle to gain recognition within an ecosystem optimized for productivity over breakthrough potential (Bhattacharya & Packalen, 2020).

To address this gap, the Disruption Index (D-index)—originally developed to measure technological innovation (Funk & Owen-Smith, 2017)—was introduced as the first metric to capture scientific breakthroughs (Wu et al., 2019), offering an alternative to traditional citation counts. The D-index has gained wide interest among scholars, funding agencies, media, and the general public due to its success in highlighting landmark discoveries. For example, it assigns a high disruption score to Watson and Crick's 1953 DNA paper (D = 0.96, top 1%) and a low score to the 1999 Human Genome Project paper (D = -0.017, bottom 6%), even though both have similarly high citation counts. This distinction, invisible to citation metrics alone, underscores the D-index's ability to differentiate truly novel work from cumulative efforts.

Despite widespread interest, efforts to expand and replicate the D-index have been hindered by limited theoretical clarity, lack of large-scale open data, and inconsistent implementation. For example, a study of historical D-index trends using Web of Science data reported a consistent decline over the past six decades (Park et al., 2023). Replication attempts using alternative data sources—without access to Web of Science—found that the decline is much smaller after removing zero-reference works, raising concerns about whether the original finding was an artifact (Holst et al., 2024). However, the dataset Holst et al. used (Z. Lin et al., 2023) contained roughly three times more zero-reference records than the original study, as it combined journal articles—which typically include references—with other scholarly works, such as book chapters, which often do not. Recent analyses confirm that Park et al.'s findings remain robust even after excluding zero-reference items (Park et al., 2025).

A second example concerns the relationship between team size and the D-index. Using Web of Science data, Wu et al. (2019) reported a consistent negative association between team size and disruption. In contrast, Petersen, Arroyave, and Pammolli (2024, 2025), using Microsoft Academic data, found a positive marginal effect after controlling for covariates, raising the possibility that the earlier findings were influenced by untested confounders. However, while Wu et al. used the longest available time window in calculating the D-index, the replication relied on a fixed 5-year window, which may bias results in favor of large teams. Since small teams often take longer to accumulate citations—acting as "sleeping beauties"—short windows can understate their disruptive impact. Reproducing Petersen et al.'s results using the same model, data, and software, Wu et al. confirmed the positive effect under a 5-year window, but showed that the negative association reemerges when extending the window, with a clear turning point at 10 years (Lin et al., 2025).

Without a clear understanding of its theoretical foundations or standardized practices for datasets and code, the D-index risks remaining a fun intellectual toy model rather than a practical tool for research evaluation and policy-making. This raises a central question: what does the D-index actually measure?

In this paper, we position the D-index as a foundational metric for identifying breakthrough research. Contrary to the common belief that it measures absolute innovation (Park et al., 2023; Wu et al., 2019), we show that the D-index captures *relative* innovation—a paper's ability to displace its most-cited reference. In doing so, it reflects scientific progress as the replacement of older answers with better ones to the same fundamental question—much like light bulbs replacing candles. We support this insight through mathematical analysis, expert surveys, and large-scale bibliometric evidence. To facilitate replication, verification, and broader use, we release a dataset of D-index values for 49 million journal articles (1800–2024) based on OpenAlex.

## 2. The Disruption Index Captures Displacement Between a Paper and its Most Cited Reference

Calculating the D-index begins by classifying all subsequent papers citing a focal paper *p* into three types: those that cite *p* but not its references (type *i*), those that cite both *p* and its references (type *j*), and those that cite only its references but not *p* (type *k*). The D-index is then defined as the difference between the number of type *i* and type *j* papers, normalized by the total number of type *i*, *j*, and *k* papers (Funk & Owen-Smith, 2017; Wu et al., 2019). This formulation captures the extent to which a focal paper displaces its predecessors in subsequent literature. The formula is shown in Eq. 1, with *N* denoting the number of each type of paper. A simplified illustration is provided in Fig. 1.

$$D_p = \frac{N_i - N_j}{N_i + N_j + N_k} \quad \text{Eq. (1)}$$

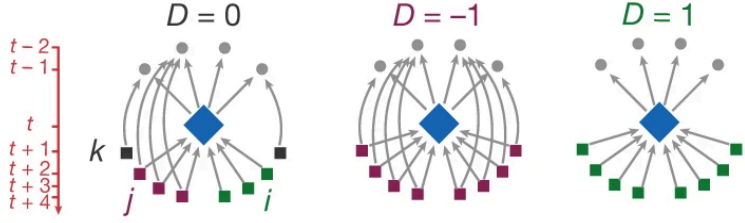

**Figure 1. Simplified illustration of *D*-index.** The figure is reproduced from our earlier research (Wu et al., 2019).

Through years of investigation, we find that the D-index captures displacement between a paper and its most cited reference. This is because to achieve a high D-index, a paper must compete against its references for future citations, primarily against the most cited reference due to the long-tail distribution of citation impacts among these references. We can see this insight by rearranging Eq. 1:

$$D_p = \frac{(N_i - N_j)/(N_i + N_j)}{1 + N_k/(N_i + N_j)} \approx \frac{1}{1+R_k} d_p \quad \text{Eq. (2)}$$

Here, we decompose the D-index into two terms, $d_p$ and $R_k$. $d_p = (N_i-N_j)/(N_i+N_j)$ is a "local" measure reflecting the focal paper *p*'s intrinsic innovative level based on the two types of citing papers of it. The other term, $R_k = N_k/(N_i+N_j)$, approximates $C_{total}/(N_i+N_j)$, where $N_k$ (the exclusive citations to the references) serves as a proxy for the total citations to the references $C_{total}$. This approximation holds because shared citations between the focal paper and its references are generally an order of magnitude smaller than total reference citations. As a result, $R_k$ quantifies the ratio of citation impact between the focal paper's references and the focal paper itself.

Notably, this competition primarily occurs between the focal paper and its most cited reference, as reference citations follow Zipf's law:

$$C_r \propto c \frac{1}{(b+r)^a} \quad \text{Eq. (3)}$$

where *c* is a constant, rank *r* represents the decreasing rank of the reference by citations, and $C_r$ denotes the citation impact of the corresponding reference. Parameter *a* indicates how unequal the citation distribution is, and parameter *b* is a fitted constant. Empirical results are shown in Fig. 2.

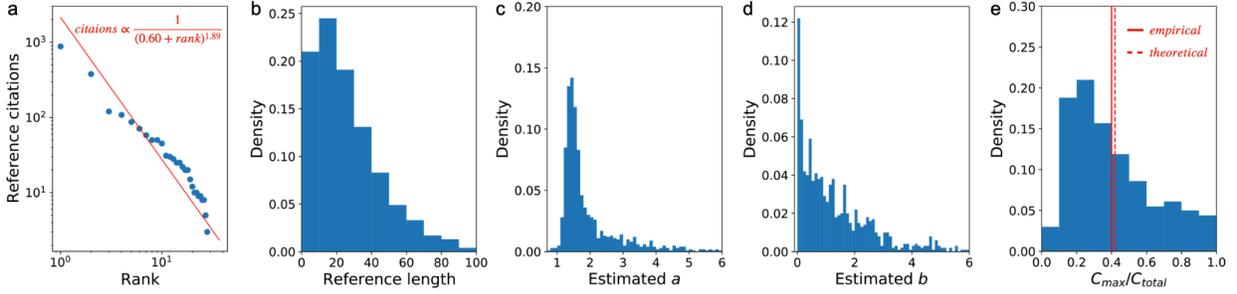

**Figure 2. Zipf's law of reference citation impacts.** We randomly selected 1,000 OpenAlex journal papers from 1900 to 2020 with three or more references. This reference length threshold ensures accurate parameter estimation. (**a**) Citation impact of a paper's references plotted by decreasing rank (blue dots), with a Zipf's law fit overlaid (red line). The estimated parameters are $a = 1.89$ and $b = 0.60$. (**b**) The distribution of the reference lengths, with an average of 29. (**c**) The distribution of the estimated $a$ values for the 1,000 papers; 99.3% of them are all greater than 1, with an average of 2. (**d**) The distribution of the estimated b values, with an average of 1.4. (**e**) The distribution of empirical values of $C_{max}/C_{total}$, with an average of 0.40 (red solid line), closely matching the theoretical prediction of 0.42 (red dashed line).

To quantify the dominance of the most-cited reference ($r = 1$) over the total citations to all references, we calculate the following ratio:

$$C_{max}/C_{total} = c\frac{1}{(b+1)^a} / \sum_{r=1}^{N} c\frac{1}{(b+r)^a} \qquad \text{Eq. (4)}$$

where $C_{max}$ represents the citation count of the most cited reference, and $C_{total}$ is the total number of citations across all references. Note that $C_{total}$ can also be expressed as:

$$C_{total} = \sum_{r=1}^{N} c\frac{1}{(b+r)^a} \approx c \int_{1}^{N} \frac{dx}{(b+x)^a} \approx c\left(\frac{(b+N)^{1-a}}{1-a} - \frac{(b+1)^{1-a}}{1-a}\right) \qquad \text{Eq. (5)}$$

Thus, the ratio $C_{max}/C_{total}$ simplifies to:

$$C_{max}/C_{total} \approx \frac{1}{(b+1)^a} / \left(\frac{(b+N)^{1-a}}{1-a} - \frac{(b+1)^{1-a}}{1-a}\right) \approx \frac{1}{(b+1)^a} / \frac{(b+1)^{1-a}}{a-1} \approx \frac{a-1}{1+b} \qquad \text{Eq. (6)}$$

This is because $a > 1$ (Fig. 2c), so the exponent $(1-a)$ is negative, making $(b+N)^{1-a}$ a rapidly shrinking term as $N$ increases. As a result, its influence is negligible. Substituting the empirical values $a = 2.0$ and $b = 1.4$ into Eq. 6, we estimate $C_{max}/C_{total} \approx 0.42$, closely matching the empirical value of 0.40 and validating our analytical reasoning (Fig. 2e).

Given that $C_{total} \approx 2.5 C_{max}$ we can rewrite Eq. 2 as:

$$D_p = \frac{1}{1+R_k}d_p \approx \frac{1}{1+C_{total}/(N_i+N_j)}d_p \approx \frac{1}{1+2.5*C_{max}/(N_i+N_j)}d_p = \frac{1}{1+2.5*C_{max}/C_p}d_p \qquad \text{Eq. (7)}$$

where $C_p = N_i + N_j$, or,

$$D_p \approx \frac{1}{1+b_p} d_p \qquad \text{Eq. (8)}$$

Where $b_p = C_{max}/C_p$. Eq. 8 shows that the D-index is primarily determined by two variables, $d_p$ and $b_p$. We define these terms, discuss their interpretations, and present their empirical values below.

**The local displacement factor,** $d_p = N_i/C_p - N_j/C_p = p_i - p_j$, measures the disparity between the probability of two types of citing papers, those that cite only the focal paper $p$ and those that cite it along with its references. This metric reflects the intrinsic innovativeness of the focal paper. If $d_p > 0$, the focal paper tends to undermine the influence of prior work; if $d_f < 0$, it consolidates and enhances previous contributions; and if $d_f = 0$, it is neutral. In our dataset, only 38% of papers are disruptive, with a median $d_p$ of –0.2 (Fig. 3a).

**The knowledge burden factor,** $b_p = C_{max}/C_p$, represents the ratio of the most-cited reference's citation impact to that of the focal paper. The name draws from the "burden of knowledge" theory (Jones, 2009) and captures the challenge a paper faces in displacing or consolidating its most influential predecessor. A truly disruptive paper must surpass the impact of its most-cited reference—thus carrying little burden ($b_p < 1$).

In our dataset, however, most papers fall short of this standard: $b_p$ has a median value of 119, yielding $1/(1+b_p) \approx 0.01$ (Fig. 3b). This helps explain why most papers have D-index values close to zero—not because they lack the potential to displace or build upon prior work, but because their influence remains largely localized within a field rather than recognized across fields. As Newton famously said, great papers may "stand on the shoulders of giants"—but not all who stand on giants become giants themselves. Only a small fraction of papers—about 1%—manage to shed the burden of past knowledge and displace their predecessors ($b_p < 1$).

Combining these two factors yields a characteristic D-index value of $-0.2 \times 0.01 = -0.002$ (Fig. 3c), helping explain why most papers have a small negative D-index. Of course, this is a rough approximation—the actual median D-index is even smaller, at approximately –0.0001. This reflects how science typically progresses: through incremental contributions by the many, with only a few papers redefining the field. Notably, the average D-index for Nobel Prize–winning papers (N = 877, 1902–2009) is just 0.1 (Wu et al., 2019).

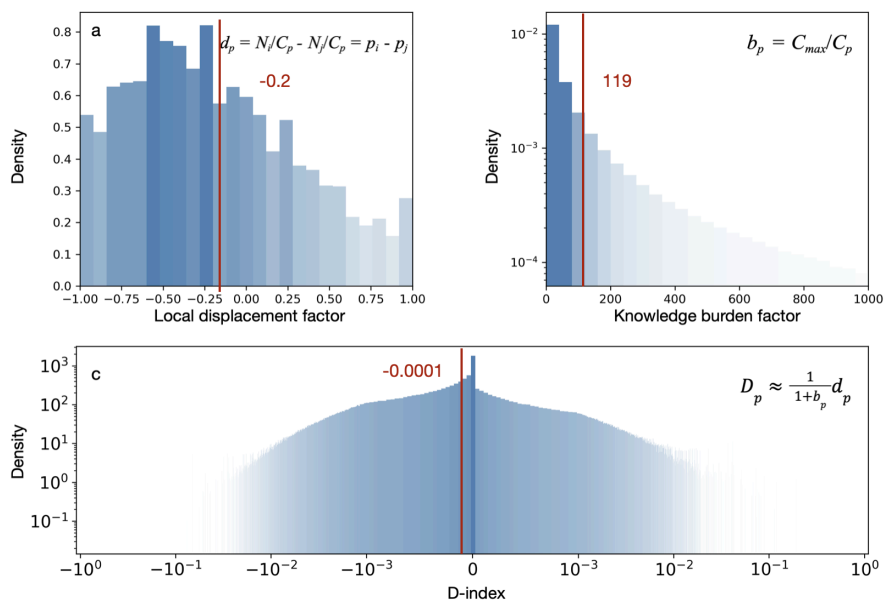

**Figure 3. Decomposing the D-index.** The distribution of the local displacement index (**a**) and the knowledge burden factor (**b**) for 22 million OpenAlex journal papers with ten or more citations. This citation threshold ensures sufficient variation in the data. With too few citations, the variables—being ratios of natural numbers—take on only a few discrete values, making it difficult to observe continuous change. Among these papers, 65% have negative $d_p$, 32% have positive $d_p$, and 3% have a $d_p = 0$. For the knowledge burden factor, 98.9% of papers have $b_p > 1$, 1% have $b_p < 1$, and 0.1% have $b_p = 1$. The dominance of negative $d_p$ (56%) and $b_p$ greater than one (98.8%) remains highly consistent when we include all papers with at least one citation and one reference. (**c**) The distribution of the D-index across 42 million OpenAlex journal articles (1900 - 2020) with one or more citations and references. 38% of these papers have a positive D-index, while 62% have zero or negative values.

## 3. The Disruption Index Captures How New Answers Replace Old Ones to the Same Scientific Question

Given the difficulty of displacing science, how do high D-index papers emerge? Do they represent true breakthroughs, displacing their most-cited reference with genuine alternatives, or can the D-index be manipulated to obscure sources and claim intellectual credit across fields?

To investigate, we re-analyzed data from our 2019 global breakthrough survey (Wu et al., 2019). In 2019, we conducted an open-ended survey on identifying breakthrough research in science, performed in person, over the telephone, or using Skype, approved by the University of Chicago Institutional Review Board (IRB18-1248). The survey asked scholars across various fields to propose papers that either disrupt or consolidate science in their fields, using the following definitions: (a) Consolidating (labeled "developmental" in the survey) papers: Extensions or improvements of previous theory, methods, or findings; (b) Disruptive papers: Punctuated advances beyond previous theory, methods, or findings.

We provided respondents with examples like the BTW model (Bak et al., 1987) and Bose-Einstein condensation (Davis et al., 1995) papers to illustrate disrupting and consolidating papers. Respondents then proposed three to ten disruptive and developing papers. Our panel included scientists from ten prominent research-intensive institutions across the United States, China, Japan, France, and Germany, with training in mathematics, physics, chemistry, biology, medicine, engineering, computer science, psychology, and economics.

Among the 20 scholars who submitted 190 responses, all nominations for the most disruptive paper aligned with our measure, and all but six for the most consolidating paper did as well. The average D-index of papers nominated as disruptive is 0.21, placing them in the top 1% of most disruptive papers. The average D-index of papers nominated as consolidating is −0.011, placing them in the bottom 13%. This analysis yielded an overall area under the curve (AUC) of 0.83, indicating strong agreement between the D-index and expert judgment.

For the current research, we re-analyzed the data to select the top nominated papers from the survey and identified their most cited references, as presented in Table 1.

**Table 1. Breakthrough Papers Nominated in a Global Expert Survey and Their Most-Cited References (highlighted in gray).**

| Year | Authors | Paper title | D |
|---|---|---|---|
| 1953 | Watson & Crick | Molecular structure of nucleic acids: A Structure for Deoxyribose Nucleic Acid | 0.96 |
| 1953 | Pauling & Corey | A proposed structure for the nucleic acids | |
| 1967 | Mandelbrot | How long is the coast of Britain? Statistical self-similarity and fractional dimension | 0.95 |
| 1954 | Steinhaus | Length, shape and area | |
| 1963 | Lorenz | Deterministic nonperiodic flow | 0.81 |
| 1916 | Rayleigh | On convection currents in a horizontal layer of fluid, when the higher temperature is on the… | |
| 1937 | Turing | On computable numbers, with an application to the Entscheidungs problem | 0.71 |
| 1931 | Gödel | On formally undecidable theorems of Principia Mathematica and related systems I | |
| 1998 | Watts & Strogatz | Collective dynamics of 'small-world' networks | 0.53 |
| 1967 | Milgram | The small world problem | |
| 2003 | Blei et al. | Latent dirichlet allocation | 0.42 |
| 2000 | Nigam et al. | Text classification from labeled and unlabeled documents using EM | |
| 1992 | Wynn | Addition and subtraction by human infants | 0.36 |
| 1980 | Starkey & Cooper | Perception of numbers by human infants | |
| 1983 | Kirkpatrick et al. | Optimization by simulated annealing | 0.29 |
| 1953 | Metropolis et al. | Equation of state calculations by fast computing machines | |
| 1951 | Nash | Non-cooperative games | 0.28 |
| 1944 | Von Neumann & Morgenstern | Theory of games and economic behavior | |

**Table 2. Breakthrough Papers Selected for Nature's 150th Anniversary and Their Most-Cited References (highlighted in gray).**

| Year | Authors | Paper title | D |
|------|---------|-------------|---|
| 1953 | Watson & Crick | Molecular structure of nucleic acids: A Structure for Deoxyribose Nucleic Acid | 0.96 |
| 1953 | Pauling & Corey | A proposed structure for the nucleic acids | |
| 1985 | Kroto et al. | C60: Buckminsterfullerene | 0.80 |
| 1984 | Rohlfing et al. | Production and characterization of supersonic carbon cluster beams | |
| 1985 | Farman et al. | Large losses of total ozone in Antarctica reveal seasonal ClOx/NOx interaction | 0.76 |
| 1978 | Dunkerton | On the mean meridional mass motions of the stratosphere and mesosphere | |
| 1975 | Köhler & Milstein | Continuous cultures of fused cells secreting antibody of predefined specificity | 0.69 |
| 1963 | Jerne & Nordin | Plaque formation in agar by single antibody-producing cells | |
| 1976 | Neher & Sakmann | Single-channel currents recorded from membrane of denervated frog muscle fibers | 0.37 |
| 1973 | Anderson & Stevens | Voltage clamp analysis of acetylcholine produced end-plate current fluctuations at frog… | |
| 1947 | Rochesterd & Butlers | Evidence for the Existence of New Unstable Elementary Particles | 0.36 |
| 1947 | Lattes et al. | Observations on the tracks of slow mesons in photographic emulsions | |
| 1992 | Kresge et al. | Ordered mesoporous molecular sieves synthesized by a liquid-crystal template mechanism | 0.21 |
| 1967 | Gregg et al. | Adsorption surface area and porosity | |
| 1995 | Mayor & Queloz | A Jupiter-mass companion to a solar-type star | 0.21 |
| 1991 | Duquennoy & Mayor | Multiplicity among solar-type stars in the solar neighbourhood | |
| 1958 | Gurdon et.al | Sexually Mature Individuals of Xenopus laevis from the Transplantation of Single Somatic Nuclei | 0.08 |
| 1956 | King & Briggs | Serial transplantation of embryonic nuclei | |

In addition to the nominated papers from the global expert survey, we analyze a second set of extraordinary papers selected by *Nature* editors to celebrate the journal's 150th anniversary (*10 Extraordinary Nature Papers*, 2019), in order to enhance the representativeness of our dataset in capturing groundbreaking science. Notably, Watson and Crick's 1953 paper on the structure of DNA appears in both sets, reflecting broad consensus on its breakthrough significance. This set of *Nature* breakthrough papers is also highly disruptive, confirming the strong alignment between the D-index and expert judgment.

Our review of these two sets of breakthrough papers and their most-cited references reveals a consistent pattern: displacement within the same fundamental question, with newer work offering clearer or more powerful answers. In biology, for example, Watson and Crick's 1953 paper (Watson & Crick, 1953) displaced Pauling and Corey's competing model (Pauling & Corey, 1953) by introducing the double-helix structure of DNA—correcting the earlier triple-helix hypothesis and transforming molecular biology. In computer science, Turing's 1937 work (Turing, 1937) reframed Gödel's incompleteness theorems (Gödel, 1931) by introducing the Turing machine, laying the foundation for modern computation.

To examine whether the insight from case studies holds at scale, we quantified how often displacing papers ($D > 0$) and their most-cited references share overlapping topics. OpenAlex journal articles are classified into an average of two fields (e.g., Discrete Mathematics, Molecular Biology, and Organic Chemistry) using a 292-category taxonomy (Sinha et al., 2015). We analyzed 49,077 high-impact (>100 citations) and highly disruptive ($D>0.2$) papers (1900–2020) alongside their most-cited references. If displacement were random, combinatorial calculations predict a 0.014 probability that these highly disruptive papers share a field with their most-cited reference (Eq. 9). Yet, empirical analysis shows a 0.52 probability—37 times higher than expected. The high topic alignment between displacing papers and their top references suggests that breakthroughs often occur as purposeful innovations, rather than merely from the random recombination of prior knowledge (Weitzman, 1998).

$$p = 1 - \frac{\binom{290}{2}}{\binom{292}{2}} \approx 0.014 \qquad \text{Eq. (9)}$$

**4. Evaluating the Robustness of the Disruption Index**

Our understanding of what the D-index actually measures—the displacing relationship between a focal paper and its most-cited reference—sheds light on ongoing debates about its technical complexity (Bentley et al., 2023; Holst et al., 2024; Leydesdorff et al., 2021; Macher et al., 2023), as well as its patterns and interpretation (Park et al., 2023). Below, we address recent concerns aimed at informing better decisions on these technical issues.

**4. 1 The D-index and Reference Length**

As science advances, more papers are published every year, and each paper cites more prior papers. This phenomenon is called "citation inflation" due to its similarity with monetary inflation caused by an increase in the money supply in economic systems (Pan et al., 2018). A recent study raised concerns about whether this could confound the observed decline in the average value of the D-index of all papers (Park et al., 2023), and make the temporal analysis of the D-index challenging in general (Petersen et al., 2024, 2025). The rationale is that the more references a focal paper includes, the less likely it is to have a high D-index, as it becomes increasingly difficult to eclipse all its references. If this reasoning holds, the D-index may converge toward zero as citation counts continue to inflate.

While this rationale helps highlight the growing knowledge burden faced by papers over time, it is not directly relevant to understanding the D-index. As we have shown, the true burden—captured by the knowledge burden factor $b_p$—is not about the number of references but about the impact of the most-cited reference. In other words, the challenge lies not in how many prior works exist, but in which "giant" in the canonical literature the focal paper attempts to displace—not the era in which the paper is published. To better illustrate this point, we show that the D-index is independent of reference length after accounting for the local displacement index ($d_p$) and the burden factor ($b_p$). See Fig. 4.

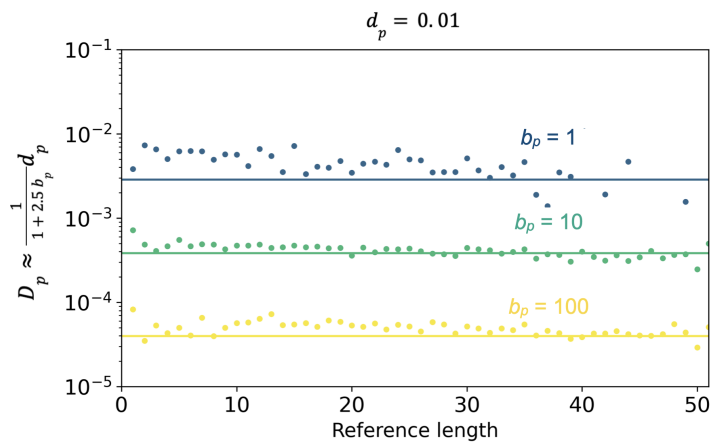

**Figure 4. D-index is independent of reference length after accounting for $d_f$ and $b_f$.** We select 929,900 papers with an average local displacing index ($d_p$) of 0.01 and ten or more citations, with values ranging from 0 to 0.05. We then further divided them into subgroups based on the burden factor, including $b_p$=1 (N = 1,022), $b_p$=10 (N = 15,682 papers), and $b_p$=100 (N = 16,706). The empirical values of the D-index for these papers align with their theoretical predictions.

We acknowledge that the analysis here is based on approximations. In research evaluation, when assessing the overall innovation performance of a collection of papers, focusing on the sign of the D-index rather than its average value can help minimize the influence of reference length (Petersen et al., 2024).

### 4. 2 The D-index and Citation Window Length.

The D-index has a life cycle: it changes as the focal paper and its references receive more citations and stabilizes when citation growth ceases (Bornmann & Tekles, 2019; Lin et al., 2022). Therefore, to calculate a stabilized D-index, the citation window, i.e., the time window for analyzing subsequent citations to the focal paper, must not be too short. Recent studies have used a five-year citation window (Park et al., 2023; Petersen et al., 2024, 2025), but we do not recommend this approach, as the D-index may take ten years or more to fully stabilize—especially for disruptive papers, which tend to accumulate recognition more slowly than consolidating ones (Bornmann & Tekles, 2019; Lin et al., 2022).

The short, five-year citation window has caused issues; for example, recent research observed the positive marginal effect of team size on the D-index while accounting for various confounders (Petersen et al., 2024, 2025), contrary to previous reports (Wu et al., 2019). This is because it takes a longer time for small teams to accumulate citations compared to large teams. As a result, a short time window underestimates the D-index of small teams (see Extended Data Fig. 7 in previous research (Wu et al., 2019)). Here we show in Fig. 5 that with an extended citation window, the negative effect of team size on the D-index reported in (Wu et al., 2019) is recovered using the same model suggested by (Petersen et al., 2024).

$$D_p = b_0 + b_k \ln k_p + b_r \ln r_p + b_c \ln c_p + D_t + \epsilon_t \qquad \text{Eq. (10)}$$

Eq. 10 controls for the temporal change of the D-index using yearly fixed effects, denoted by $D_t$. The results of the ordinary least squares (OLS) estimation, conducted using the STATA 13.0 package *reg*, are shown in Fig. 5. These results are based on a dataset of 1.7 million papers with $1 \leq k_f \leq 10$ coauthors, $5 \leq r_f \leq 50$ references, and $10 \leq c_f \leq 1000$ citations, following the same parameters as in (Petersen et al., 2024, 2025). The independent variables are modeled using a logarithmic transform due to their right-skewed distribution. The marginal effects of team size on the D-index are calculated with all other covariates held at their mean values.

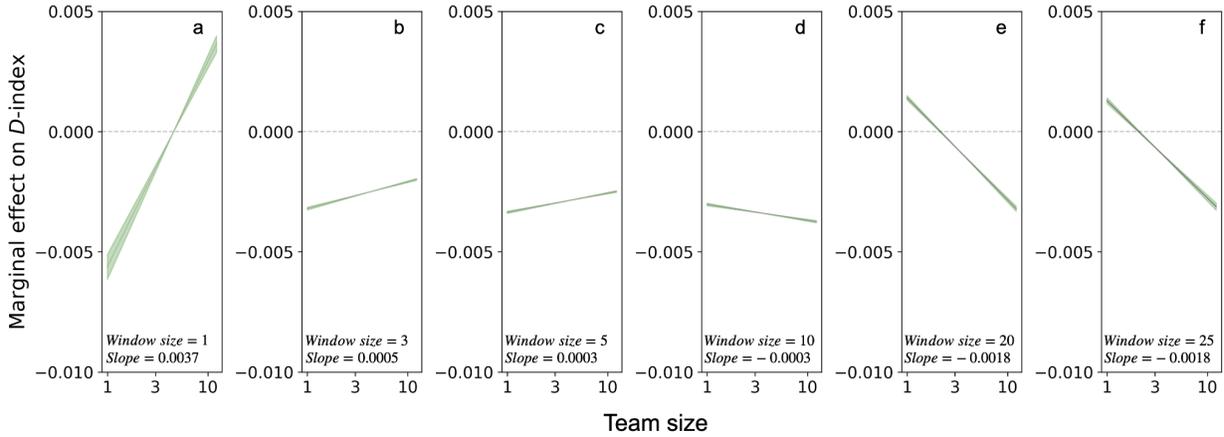

**Figure 5. The negative impact of team size on the D-index is recovered with long-term citations.** To examine how citation window length moderates the relationship between team size and the D-index, we analyzed six annual cohorts of papers, each receiving citations from subsequent papers published through 2020. Our dataset includes 47,129 papers from 2019, 271,496 from 2017, 444,675 from 2015, 536,463 from 2010, 344,582 from 2000, and 226,358 from 1995, corresponding to citation windows of 1, 3, 5, 10, 20, and 25 years, respectively. The regression coefficients (slopes) estimated from Eq. 10 are presented, with marginal effects calculated while holding all other covariates at their mean values. Light green confidence intervals are shown around the regression lines. This figure is reproduced from our early research, where we also controlled for cohort effects and found that the time window effect remains consistent (Lin et al., 2025).

### 4. 3 The D-index's Discriminative Power and Emerging Alternatives

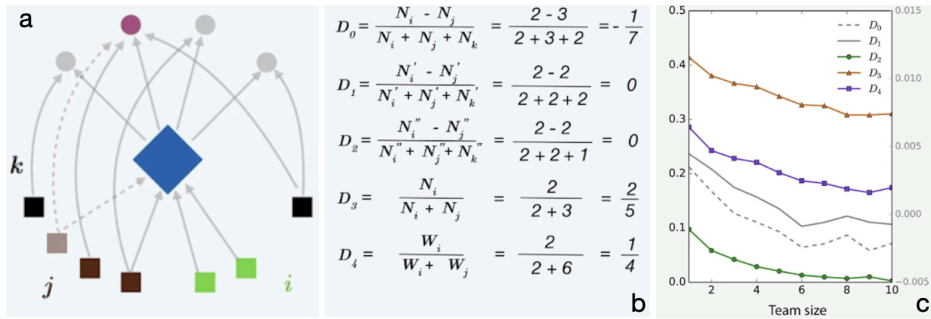

**Figure 6. Alternative versions of *D*-index.** The figure is reproduced from our earlier research (Wu et al., 2019). (**a**) A simplified citation network comprising focal papers (diamonds), references (circles), and subsequent work (rectangles). Subsequent work may cite (1) only the focal work (*i*, green), (2) only its references (*k*, black), or (3) both focal work and references (*j*, brown). A reference identified as popular is colored in red, and self-citations are shown by dashed lines (with corresponding subsequent work colored in light brown). (**b**) Five definitions of the D-index are provided for comparison. $D_0$ is the definition used in the main text. $D_1$ is defined the same way as $D_0$, but with self-citations excluded. $D_2$ is defined the same way as $D_0$ but only considers popular references. In the empirical analysis, we identified references as popular that received citations within the top quartile of the total citation distribution (≥24 citations). $D_3$ simplifies $D_0$ by only measuring the fraction of papers that cite the focal paper and not its references, among all papers citing the focal paper. $D_4$ is similar to $D_3$ but weighted by the number of citations. For example, if a single referenced paper is cited five times, then it receives a count of five rather than one. (**c**) All alternative measures to the D-index decrease consistently with team size. $D_0$ and $D_1$ are indexed by the right y-axis and other disruption measures are indexed by the left y-axis. 100,000 randomly selected Web of Science papers (97,188 papers remained after excluding missing data) are used to calculate these values.

Recent research has raised concerns about the D-index's discriminative power, particularly because its numerator is bounded while its denominator is unbounded (Eq. 1). This formulation tends to produce values close to zero, potentially undermining its discriminative power (Petersen et al., 2024). To address this issue, some studies have explored alternative versions of the D-index (Bornmann et al., 2020). The analysis in the previous sections has clarified why the D-index is typically small, and we now highlight the unique value of the original D-index for two key reasons.

$$D_p \approx \frac{1}{1+b_p} d_p \qquad \text{Eq. (8)}$$

First, the original D-index is highly effective in identifying revolutionary work. Our decomposition of the D-index into a local displacement factor ($d_p$) and a knowledge burden factor ($b_p$) explains why most D-index values cluster near zero: the majority of papers cite canonical literature with much higher citation impact. While some scholars view this as a limitation, it actually highlights a small subset of papers ($b_p < 1$, less than 1%) whose role in displacing or consolidating their most-cited reference is substantial. These papers stand apart from the vast majority, whose influence remains minimal in comparison—regardless of whether their contribution is disruptive or consolidating (as indicated by the sign of $d_p$).

For example, given a local displacement index $d_p = 0.5$, maintaining this displacing effect globally is increasingly difficult. If the focal paper has the same citation impact as its top reference ($b_p=1$), which is already rare, $D_f$ is reduced to 0.5/2= 0.25 (see Eq. 8). If the focal paper has twice the citation impact of its top reference ($b_p=0.5$), $D_p$ reduces to 0.5/1.5=0.33. Only when the focal paper has ten times the citation impact of its top reference ($b_p=0.1$), $D_p$ is largely preserved: 0.5/1.1=0.45. However, such occurrences are extremely rare and may only happen a couple of times in a decade (0.03%), especially if the top reference is canonical literature. In other words, when a highly positive D-index is observed, it means the idea presented by the paper not only substitutes its top reference but is also well-recognized by the field and beyond.

Therefore, if the goal is to identify paradigm-shifting work in the history of science, the original D-index ($D_p$) has a higher discriminative power. If the goal is only to identify different contributions within the scope of normal science, the local displacement index ($d_p$) is more effective.

Second, alternative versions of the D-index exhibit similar behaviors. For example, a recent study stated that "the results of a factor analysis show that the different variants measure similar dimensions" (Bornmann et al., 2020). Our previous research (Wu et al., 2019) has also considered five versions of the D-index, all of which demonstrated consistent correlation with another variable, team size (Fig. 6). This includes a version ($D_3$) that has excluded self-citations, which may, therefore, ease the concern raised in (Petersen et al., 2024).

Additionally, recent research proposed that the original D-index may not be accurate and should be weighted by the number of citations to account for both the magnitude and reach of high D-index papers (Bentley et al., 2023). This is similar to the weighted version ($D_4$) presented in Fig. 6.

**4.4 The D-index and Betweenness Centrality in Citation Networks**

Recent literature suggests that the *D*-index is a specific form of node centrality in citation networks: betweenness (Gebhart & Funk, 2023). Betweenness centrality measures how often a node appears on the shortest paths between other nodes, indicating its role as a bridge within the network (Freeman, 1977). While we agree with this topological interpretation, we would like to emphasize that it should not confuse the originality and meaning of the *D*-index.

First, using node centrality to measure paper importance has a long history (Price, 1965), inspiring the PageRank algorithm in information retrieval (Brin & Page, 1998). However, to our knowledge, these network measures rarely leverage the hidden time dimension as the D-index does. The D-index uniquely captures this temporal dimension, highlighting papers with high values as "gatekeepers in time" or "structural holes in time."

Second, interpreting the *D*-index as merely betweenness centrality in networks risks focusing on the strategic advantage of high *D*-index papers as "knowledge brokers" and ignoring their inherent intellectual contributions. While being a knowledge broker in social networks often reflects social capital advantages (Burt, 2004; Granovetter, 1973), attaining this role in citation networks—especially across time—is hard-earned. For example, in our Breakthrough Papers Dataset, the 1998 small-world paper by Duncan Watts and Steve Strogatz (Watts & Strogatz, 1998) displaced Stanley Milgram's 1967 paper (Milgram, 1967) (see Table 1). It is an oversimplification to assume that subsequent citations of Watts and Strogatz were simply due to ignorance of Milgram's work, especially considering that researchers are actually trained to discover and cite the original literature. Based on our interview with experts, the high *D*-index of Watts and Strogatz correctly reflects its radical advancement from Milgram's work, by providing a novel mathematical framework to quantify the small-world phenomenon beyond social networks.

**4.5 The Asymmetry of the D-index Distribution**

Regarding the interpretation of the D-index, it is important to note that papers with a negative D can also significantly contribute to science, as described by the term "consolidation," hence the name "CD-index" (Funk & Owen-Smith, 2017). For example, Wolfgang Ketterle et al.'s paper on Bose-Einstein Condensation (Davis et al., 1995), which validated the theory proposed by Albert Einstein and Satyendra Nath Bose through lab experiments, has a D=-0.58 (bottom 3% among all papers) (Wu et al., 2019). Ketterle won the Nobel Prize in Physics in 2001 for this work. Another example is the 2001 human genome paper (Lander et al., 2001), an important milestone in genomic research resulting from massive international collaboration (D = -0.017, bottom 6%). Both of these works consolidate revolutionary scientific ideas—the Einstein-Bose condensation theory and the DNA structure—rather than displacing them, yet they still represent fundamental progress in science.

The asymmetric distribution of the D-index (62% D ≤ 0) suggests that consolidating innovation is the norm in science. This pattern differs from that in technology, where more patents have a positive D-index (62%) than a negative one (38%), based on open-source data we published (Y. Lin et al., 2023). It would be interesting to explore whether this reflects fundamental differences in the level of path dependency between science and technology.

**5. Opening the D-index Data for 49 Million OpenAlex Journal Articles**

A recent study raised concerns about missing data affecting the calculation of the D-index (Holst et al., 2024). Accurate D-index calculation relies on the complete retrieval of paper references. Missing values in paper references can distort the D-index. For example, a paper with references but no citations results in a zero D-index. This can be misleading as papers with many citations but evenly split between subsequent papers that also cite its reference or disregarding them also result in D = 0. Similarly, papers with citations but no references may misleadingly result in a zero D-index. This can distort the explanation of results, as our previous analysis shows that achieving a high positive D-index is very difficult and a rare event in the real world.

These issues can affect the D-index itself and skew downstream analyses correlating D with other variables, especially if missing data is unevenly distributed. To ensure the D-index is as reliable as possible, we only include papers with one or more references and citations in our analysis in this manuscript and previous studies (Lin et al., 2022, 2023; Wu et al., 2019). After all, the D-index measures intellectual contribution based on citation practices, reflecting how a focal work relates to preceding major ideas as determined by the subsequent papers. Without reference or citation data, this analysis is meaningless.

We also recommend focusing on one type of scholarly work at a time in calculating the D-index rather than mixing journals, conferences, theses, books, or essays, which have different citation practices and could affect D-index interpretation as in (Z. Lin et al., 2023). In our previous studies, we typically focused on journal articles to minimize issues from varying citation practices. This is because citation practices are more established for academic journals, and the peer-review mechanism further serves as quality control for these norms.

Indeed, without an open and scalable infrastructure, the D-index risks remaining an intellectual toy model rather than a practical tool for research evaluation and policy-making. To bridge this gap and facilitate replication, validation, and broader use, we release a dataset of D-index values for 49 million journal articles (1800–2024) based on OpenAlex. The dataset is available at [https://dataverse.harvard.edu/dataverse/OpenAlex_D_index], along with the code and workflow used to generate it from the raw OpenAlex data.

## 6. Conclusion and Discussion

In conclusion, we suggest that the D-index measures how a new idea displaces older ones while addressing similar questions or phenomena (Small, 1978). Rather than assessing the inventive level of a single paper in isolation, the D-index captures the displacement relationship between a focal paper and its most-cited reference. This perspective emphasizes the continuity and evolution of scientific knowledge, highlighting historical trajectories rather than sudden ruptures. While "Disruption" remains a compelling name, it may be more accurate to interpret the D-index as a *Displacement Index*. We hope this clarification improves understanding of the metric and promotes its more appropriate use in research evaluation (Leibel & Bornmann, 2024).

This study, along with the released D-index dataset, has the potential for a significant and lasting impact on research evaluation, funding policy, and open science. First, it empowers scholars, institutions, and media worldwide with a free and accessible dataset for identifying breakthrough papers. Second, it supports the Science of Science community by providing an open, regularly updated D-index dataset based on OpenAlex—the largest open scholarly database globally—ensuring a robust and sustainable infrastructure for evaluating innovation. Third, it informs the design of new funding mechanisms by collaborating with agencies and policymakers to translate D-index insights into practical evaluation strategies, starting with discussions at our planned policy-focused workshop. Together, these efforts promote transparency, accessibility, and a deeper understanding of breakthrough science.

**Acknowledgments.** We are grateful for support from the National Science Foundation grant SOS: DCI 2239418 (L.W.).